\documentclass[aps,onecolumn,nofootinbib,superscriptaddress,floatfix]{revtex4}

\usepackage{color,amsmath,amssymb,graphicx,latexsym,subfigure}

\usepackage{multirow}
\usepackage{ulem}
\usepackage{subcaption}

\usepackage{float}
\usepackage{booktabs}
\usepackage{threeparttable}
\usepackage[colorlinks,linkcolor=green, anchorcolor=green,citecolor=green]{hyperref}
\usepackage{ulem,soul}
\usepackage{xcolor}

\newcommand{\be}{\begin{equation}}
	\newcommand{\ee}{\end{equation}}
\newcommand{\ba}{\begin{eqnarray}}
	\newcommand{\ea}{\end{eqnarray}}

\newcommand{\gsim}{\mathrel{\hbox{\rlap{\lower.55ex \hbox {$\sim$}}
			\kern-.3em \raise.4ex \hbox{$>$}}}}
\newcommand{\lsim}{\mathrel{\hbox{\rlap{\lower.55ex \hbox {$\sim$}}
			\kern-.3em \raise.4ex \hbox{$<$}}}}

\makeatletter
\def\section{\@startsection {section}{1}{\z@}%
	{-3.5ex \@plus -1ex \@minus -.2ex}%
	{2.3ex \@plus.2ex}%
	{\normalfont\bfseries\boldmath\rightskip\z@}}
\makeatother

\hypersetup{colorlinks=true,
	breaklinks=true,
	pdfstartview=Fit,
	linkcolor=blue,
	citecolor=blue,
	urlcolor=blue}

\begin{document}
\hfill  USTC-ICTS/PCFT-26-06

\title{Realization of quintom dark energy after DESI DR2 in Nieh-Yan modified teleparallel gravity}

\author{Yuxuan Kang}
\email{yxkang@mail.ustc.edu.cn}
\author{Mingzhe Li}
\email{limz@ustc.edu.cn}
\author{Changzhi Yi}
\email{ycz20010527@mail.ustc.edu.cn}
\affiliation{Interdisciplinary Center for Theoretical Study, University of Science and Technology of China, Hefei, Anhui 230026, China}
\affiliation{Peng Huanwu Center for Fundamental Theory, Hefei, Anhui 230026, China}

\begin{abstract}
Recent observations from the DESI Collaboration indicate a preference for quintom dark energy, i.e., its equation of state evolves across the cosmological constant boundary $w=-1$. It is well known that models with single perfect fluid or single scalar field minimally coupled to Einstein gravity develop perturbative instabilities around the crossing, thereby cannot realize the quintom scenario. In this paper, we propose a method to circumvent the instability problem of these models by considering the coupling of dark energy to the Nieh-Yan density within the framework of teleparallel gravity. We show that with this coupling the background evolution is not affected, but the dark energy perturbation is removed from the menu of dynamical degrees of freedom, thus avoiding the inherent difficulties in the old models. Furthermore, the Nieh-Yan coupling causes parity violation in gravitational waves, and this can be considered as a clear prediction of this mechanism. 
\end{abstract}

\maketitle

\section{Introduction}

In 1998, measurements of distances from high-redshift Type Ia supernovae (SNe) provided the first direct evidence that the Universe is undergoing accelerated expansion \cite{SupernovaSearchTeam:1998fmf,SupernovaCosmologyProject:1998vns}. The simplest explanation for these observations is that they are due to a new form of matter with negative equation of state (EoS) called dark energy. Traditionally, this role has been played by a small positive cosmological constant $\Lambda$, whose EoS is $w=-1$ at all times. More recently, however, observations have pointed toward dynamical dark energy as the driver of the accelerated expansion. For example, the combination of Planck cosmic microwave background (CMB) data with weak lensing measurements from the Canada–France–Hawaii Telescope Lensing Survey (CFHTLenS) reveals a preference, at the 2$\sigma$ level, for a dynamical dark energy EoS relative to the $\Lambda$CDM model \cite{Planck:2015fie,Planck:2018vyg}. Analyses combining SNe datasets with baryon acoustic oscillation (BAO) measurements from the Dark Energy Spectroscopic Instrument (DESI) imply dynamical dark energy at a confidence level of 2.5-3.9$\sigma$ \cite{DESI:2024mwx}. Furthermore, these results exhibit a preference for a quintom behavior \cite{Feng:2004ad,Cai:2025mas}, wherein its EoS crosses the cosmological constant boundary $w=-1$ in the recent past. Most recently, a joint analysis that includes the most recent DESI Data Release 2 and the SNe constraint reveals a preference at the  2.8-3.8$\sigma$ \cite{DESI:2025zgx,DESI:2025fii,DESI:2025wyn,DES:2025sig}.

Typical dynamical dark energy models, such as quintessence \cite{Ratra:1987rm,Wetterich:1987fm,Caldwell:1997ii} and phantom \cite{Caldwell:1999ew} models, cannot cross the cosmological constant boundary. The most important property of quintom dark energy \cite{Feng:2004ad} is that its EoS crosses $-1$ during evolution. According to the direction of transition, the quintom dark energy can be classified as type A and type B models. For quintom-A, its EoS transited from the phase of $w>-1$ to that of $w<-1$ as universe expanding, while quintom-B has the opposite behavior. It is worth pointing out that current data, as shown in   \cite{DESI:2025zgx,DESI:2025fii,DESI:2025wyn} and other analyses, favor the quintom-B scenario. This challenges the dark energy model building. Simple models based on minimally coupled perfect fluid or kessence-like field \cite{Chiba:1999ka,Armendariz-Picon:2000nqq} with single component cannot realize the quintom scenario, because around the time of $w=-1$ crossing there emerges unavoidably unphysical instabilities in the dark energy perturbation. This has been noticed in Refs.\cite{Feng:2004ad,Vikman:2004dc,Hu:2004kh,Zhao:2005vj}, and is formalized as a no-go theorem in Refs.\cite{Xia:2007km,Qiu:2010ux}. As discussed in the next section, there are two kinds of instabilities involved in dark energy perturbation. One is the gradient instability caused by a negative sound speed square, another one is the ghost instability due to a kinetic term with wrong sign. The efforts to enable a $w=-1$ crossing are made by introducing explicitly more degrees of freedom into the dark energy \cite{Feng:2004ad,Hu:2004kh,Guo:2004fq,Wei:2005nw,Mughal:2020glg,Vazquez:2023kyx} or by considering higher derivatives in the scalar field model \cite{Li:2005fm,Zhang:2006ck,Cai:2007gs}. As shown in Ref.\cite{Li:2011qfa}, quintom models without pathologies in perturbations can be constructed in terms of the scalar field with degenerate higher derivatives. In addition, dark energy models with EoS crossing $-1$ have been also discussed in some theories of modified gravity or non-minimal couplings \cite{Langlois:2017mxy,Langlois:2018jdg,Horndeski:1974wa,Nicolis:2008in,Deffayet:2009wt,Deffayet:2011gz,Bamba:2010wb,Arora:2022mlo}, and more recently in Refs.\cite{Paliathanasis:2025hjw,Li:2025owk,Nojiri:2025low,Wang:2025znm,Tsujikawa:2025wca,Chen:2025ywv,Wolf:2025acj,SanchezLopez:2025uzw,Toomey:2025yuy,Paliathanasis:2025mvy,Nojiri:2025uew,Li:2026xaz,Tsujikawa:2026xqm}

In this paper we address the theoretical challenge of realizing the quintom scenario with single perfect fluid or single kessence-like field. We propose a new method to circumvent the instability of these quintom models within the framework of teleparallel gravity \cite{Aldrovandi:2013wha,Bahamonde:2021gfp}. Teleparallel gravity provides a different picture for gravitation, in which the curvature and non-metricity tensor vanish and gravity is manifested by torsion. Model equivalent to general relativity (GR) can be constructed within this framework. In this paper we will extend this GR equivalent teleparallel gravity slightly by considering the coupling of dark energy to the Nieh–Yan density \cite{Nieh:1981ww}. This formed the Nieh-Yan modified Teleparallel Gravity (NYTG) \cite{Li:2020xjt,Li:2021wij}, which has been proposed as a consistent and healthy parity-violating gravity model. We will show that, since the Nieh-Yan density vanishes in the homogeneous and isotropic background of the universe, this coupling has no effect on the background evolution of dark energy, but it provides an additional constraint and forces the gauge-invariant dark energy perturbation to vanish. This provides a mechanism to remove the dark energy perturbation from the menu of dynamical degrees of freedom, and also to avoid the inherent difficulties in the old models. Consequently, one only needs to focus on the background evolution to construct quintom models with confidence. 
In addition, the incorporation of dark energy into the NYTG model inevitably leads to parity violation in gravitational waves, and this can be considered as unique theoretical prediction of this mechanism, further signals the dynamical nature of dark energy. 

This paper is organized as follows. In Section \ref{II}, we review the no-go behavior occurred in quintom models of single minimally coupled fluid or kessence-like scalar field. To overcome these difficulties, we introduce the NYTG model in Section \ref{III}, where dark energy is coupled to the Nieh-Yan density. The detailed studies on the fluid and scalar field dark energy models within NYTG framework are presented in sections \ref{IV} and \ref{V}, respectively. Our results demonstrate the absence of dynamical degree of freedom associated with dark energy perturbation. We list two toy models and illustrate their background evolutions, which exhibit the quintom-B behavior. Section \ref{VI} presents an additional demonstration that our mechanism also predicts parity violation in gravitational waves. Finally we conclude in Section \ref{VII}. 

From now on, we adopt the unit $c=\hbar=8\pi G=1$ and the metric signature $\{-,+,+,+\}$. Here symbol c denotes the speed of light and should not be confused with the coupling constant c appearing in the Nieh-Yan coupling term introduced below. The Levi-Civita connection and its associated covariant derivative are written as $\mathring{\Gamma}^\rho_{\phantom{1}\mu\nu}=\frac{1}{2}g^{\rho\lambda}\left(\partial_\mu g_{\lambda\nu}+\partial_\nu g_{\mu\lambda}-\partial_\lambda g_{\mu\nu}\right)$ and $\mathring{\nabla}$, which we distinguish from the spacetime affine connection $\Gamma^\rho_{\phantom{1}\mu\nu}$ and its associated covariant derivative operator $\nabla$. The Levi-Civita tensor is   $\varepsilon_{\mu\nu\rho\sigma}=\sqrt{-g}\,\epsilon_{\mu\nu\rho\sigma}$, where $\epsilon_{\mu\nu\rho\sigma}$ is a totally antisymmetric symbol with $\epsilon_{0ijk}=\epsilon_{ijk}\equiv\epsilon^{ijk}$ and $\epsilon_{0123}=-\epsilon^{0123}=1$.

\section{Review on the no-go behavior of quintom model building}\label{II}

In this section, we review the no-go behavior occurred in quintom models of minimally coupled perfect fluid or scalar field with single component.
We first consider the case of dark energy as a single perfect fluid, with minimal coupling to Einstein gravity, its action usually takes the form: \cite{Jackiw:2004nm,DeFelice:2009bx,Schutz:1970my,Schutz:1977df}:
\begin{equation}\label{sfluid}
	S_f=\int d^4x \, \sqrt{-g} \, \left[-j^\mu a_\mu-f\left(\sqrt{-j^\mu j_\mu}\right) \right].
\end{equation}
The vector $j^\mu$ represents the particle number density flux and can be expressed as $j^\mu=nU^\mu$, where $n$ is the number density and $U^{\mu}$ is the normalized four-velocity, $U^\mu U_\mu=-1$. Conversely, we have $n=\sqrt{-j^\mu j_\mu}$. The vector $a_\mu$ is defined in terms of three scalar fields as $a_\mu=\partial_\mu \phi+\alpha\partial_\mu\beta$. Varying the action (\ref{sfluid}) with respect to $\phi,~\alpha,~\beta$ and $j^\mu$ leads to the following equations:
\begin{equation}\label{eomfluid}
	\mathring{\nabla}_\mu j^{\mu}=\mathring{\nabla}_\mu \left(n U^\mu\right) =0,\quad U^\mu \partial_\mu \alpha=0,\quad U^\mu \partial_\mu \beta=0,\quad a_{\mu}=f_{,n} U_{\mu},
\end{equation} 
where $f_{,n}=d f(n)/d n$. By varying the action (\ref{sfluid}) with respect to metric  $g_{\mu\nu}$, one obtains the energy-momentum tensor for perfect fluid:
\begin{equation}
	T^{\mu\nu}=Pg^{\mu\nu}+(\rho+P)U^\mu U^\nu ,
\end{equation}
with $\rho=f(n),~P=nf_{,n}-f(n)$. Since the energy density should be positive definite, the function $f(n)$ is required to be positive at all times. In the variation $j^\mu$ and $a_\mu$ are taken to be metric independent and $j_\mu=g_{\mu\nu}j^\nu$. 
The conservation law $\mathring{\nabla}_\mu j^{\mu}=0$ in any spacetime is expanded as $U^\mu \mathring{\nabla}_\mu n+(\mathring{\nabla}_\mu U^\mu)n=0$, which implies the continuity equation for the energy density $U^\mu \mathring{\nabla}_\mu \rho+(\mathring{\nabla}_\mu U^\mu)(\rho+P)=f_{,n}[U^\mu \mathring{\nabla}_\mu n+(\mathring{\nabla}_\mu U^\mu)n]=0$. The Eulerian equation of the fluid can be obtained from the rest three equations in Eq. (\ref{eomfluid}). They together form the conservation law of the energy-momentum tensor, $\mathring{\nabla}_\mu T^{\mu\nu}=0$. 

The EoS parameter of the perfect fluid is thus expressed as:
\begin{equation}\label{wfluid}
	w_f=\frac{P}{\rho}=-1+\frac{nf_{,n}}{f(n)}.
\end{equation} 
Hence one can write
\begin{equation}\label{integral}
	\rho=f(n)\propto e^{\,\int (1+w_f)  \, d \, \mathrm{ln} n}
\end{equation}
using the expression for EoS parameter $w_f$. 

We consider the background of the universe as a spatially flat Friedmann–Robertson–Walker (FRW) spacetime with the metric of the form:
\begin{equation}\label{frwmetric}
	{ds}^2=a^2(\eta)\left(-d\eta^2+\delta_{ij}dx^idx^j\right),
\end{equation}
where $a(\eta)$ is the scale factor and $\eta$ is the conformal time. At this background there is only one independent equation for the fluid, i.e., the conservation law for the particle number, $n'+3\mathcal{H} n=0$, where prime denotes the variation with respect to conformal time and $\mathcal{H}=a'/a$. This means the number density scales as $n=n_0a^{-3}$ in the expanding universe and decreases monotonically with time, here the subscript $0$ denotes the present value and we have also used the normalization $a_0=1$. The continuity equation follows automatically, 
\begin{equation}\label{fbeq}
	\rho'+3\mathcal{H}(\rho+P)=f_{,n}\left(n'+3\mathcal{H} n\right) =0 .
\end{equation}

One of the advantages of the perfect fluid model is one can know the specific form of the theoretical model from the observational data strait forwardly. Once the dependence of EoS on the scale factor $w_f(a)$ is reconstructed from data, we can obtain the function $\rho=f(n)$ through the integration (\ref{integral}) and the relation $a=(n_0/n)^{1/3}$. 

\begin{figure}[htbp]
	\centering
	\captionsetup{justification=raggedright}
	\subfigure{\includegraphics[scale=0.75]{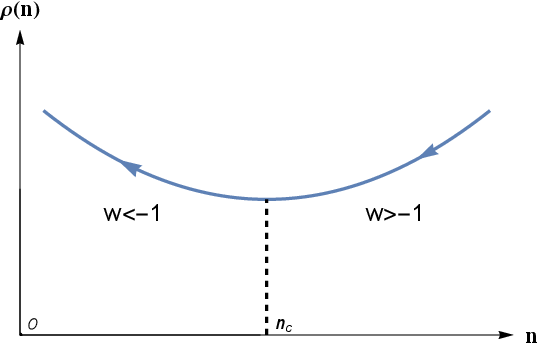}}$\quad\qquad$
	\subfigure{\includegraphics[scale=0.62]{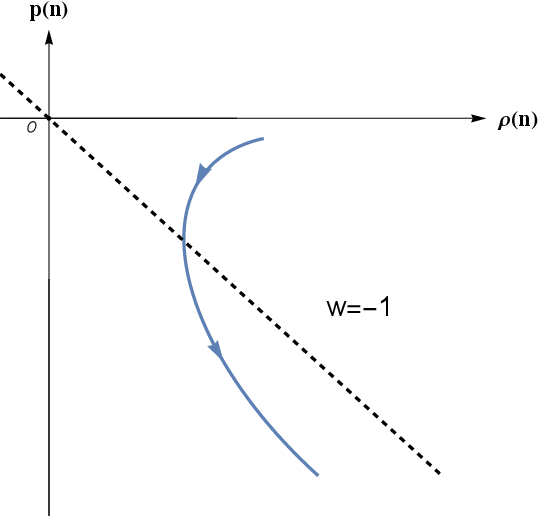}}
	\caption{\small{The left panel illustrates the energy density $\rho(n)$ as a function of the particle number density $n$ in the quintom-A scenario. The right panel shows the $P-\rho$ relation in the same scenario. Arrows indicate the direction of evolution.}}
	\label{fqa}
\end{figure}

\begin{figure}[htbp]
	\centering
	\captionsetup{justification=raggedright}
	\subfigure{\includegraphics[scale=0.75]{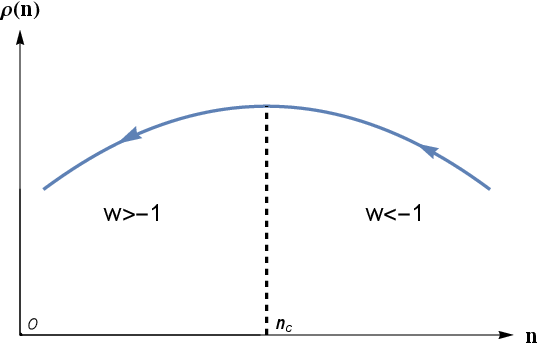}}$\quad\qquad$
	\subfigure{\includegraphics[scale=0.62]{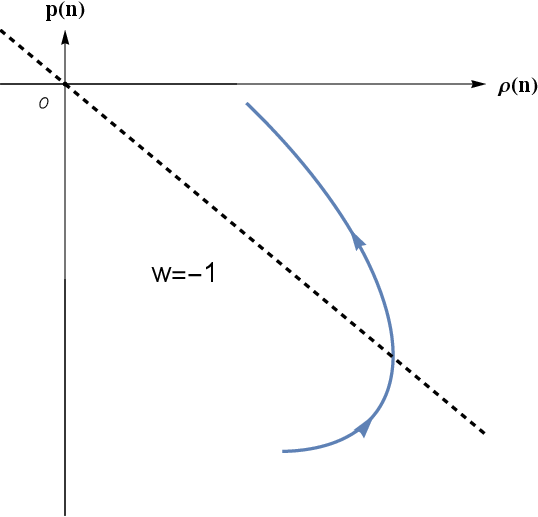}}
	\caption{\small{The left panel illustrates the energy density $\rho(n)$ as a function of the particle number density $n$ in the quintom-B scenario. The right panel shows the $P-\rho$ relation in the same scenario. Arrows indicate the direction of evolution.}}
	\label{fqb}
\end{figure}

Now we turn to the quintom behavior of the perfect fluid model. Since $\rho+P=\rho\,(1+w_f)=n f_{,n}$, it is evident that $w_f>-1$ corresponds to $f_{,n}>0$, while $w_f<-1$ corresponds to $f_{,n}<0$. Therefore, the crossing point $w_f=-1$ occurs at a local extremum $f_{,n}=0$, and a sign change in $f_{,n}$ is required to realize quintom behavior. Accordingly, quintom-A is associated with $f_{,nn}>0$, while the quintom-B implies $f_{,nn}<0$. The evolution of the energy density $\rho(n)$ and the $P-\rho$ relations for the quintom-A and quintom-B scenarios are illustrated in FIG.\ref{fqa} and FIG.\ref{fqb}, respectively. 

Now one can have a glimpse on how the pathology in dark energy perturbation arises. The $P-\rho$ relation showed that the pressure should be a multi-valued function of the energy density. Consequently, the sound speed square, defined as  $c_s^2=dP/d\rho=n f_{,nn}/f_{,n}$, diverges at the crossing point and becomes highly negative after the crossing. A negative $c_s^2$, however, will induce the gradient instability, which results in uncontrollable growth of perturbations at all scales.

To investigate the instability in more detail, we now turn to the quadratic action for the scalar perturbation of the single fluid dark energy. The total action under consideration is given by:
\begin{equation}\label{stgr}
	S=\int d^4x \, \sqrt{-g} \, \frac{\mathring{R}}{2}+S_{_{DE}}+S_m
\end{equation}
where $S_{_{DE}}$ and $S_m$ denote the actions for dark energy and matter sector, respectively. Both components share the same form as the fluid action $S_f$:
\begin{equation}\label{sfluid2}
	\begin{aligned}
		S_{_{DE}}&=\int d^4x \sqrt{-g} \left[-j_1^\mu a_{1\mu}-f_1\left(\sqrt{-j_1^\mu j_{1\mu}}\right) \right]\\
		S_m&=\int d^4x \sqrt{-g} \left[-j_2^\mu a_{2\mu}-f_2\left(\sqrt{-j_2^\mu j_{2\mu}}\right) \right]
	\end{aligned}
\end{equation}
where $a_{\mathfrak{i}\mu}=\partial_\mu\phi_\mathfrak{i}+\alpha_\mathfrak{i}\partial_\mu\beta_\mathfrak{i},~ j^\mu_\mathfrak{i}=n_\mathfrak{i}U_\mathfrak{i}^\mu$ and $\mathfrak{i}=1,2$ represent dark energy and matter sector, respectively.  The scalar perturbations of metric take the following form:
\begin{equation}\label{pertgmn}
	g_{00}=-a^2\left(1+2A\right),~g_{0i}=a^2\,\partial_iB,~g_{ij}=a^2\left[\left(1-2\psi\right)\delta_{ij}+2\partial_i\partial_jE\right].	
\end{equation}
Meanwhile, the FRW background implies $\alpha'=\beta'=0$, so that the scalar perturbations of the perfect fluid can be written as:
\begin{equation}\label{pertfluid}
	\begin{aligned}
		n(\eta,\vec{x})&=n(\eta)+\delta n\,,~~U^0(\eta,\vec{x})=\frac{1}{a}(1-A)\,,~~U^i(\eta,\vec{x})=\frac{1}{a}\,\partial_iv\,,\\
		\phi(\eta,\vec{x})&=\phi(\eta)+\delta \phi\,,\,~~\alpha(\eta,\vec{x})=\alpha(\vec{x})+\delta\alpha\,,\,~~\beta(\eta,\vec{x})=\beta(\vec{x})+\delta\beta\,,
	\end{aligned}
\end{equation}
By substituting these perturbations into the action (\ref{stgr}) and expanding to the second order, the quadratic action for scalar perturbation is obtained:
\begin{equation}\label{stgr2}
	\begin{aligned}
		S^{(2)}&=\int d^4x  \Big[z_1^2\left({\zeta_1'}^2-c_{s1}^2\partial_i\zeta_1\partial_i\zeta_1+M_1^2\zeta_1^2\right)+z_2^2\left({\zeta_2'}^2-c_{s2}^2\partial_i\zeta_2\partial_i\zeta_2+M_2^2\zeta_2^2\right)\\
		&+M_3\zeta_1\zeta_2+M_4\zeta_1\zeta_2'+M_5\zeta_1'\zeta_2\Big] ,
	\end{aligned}
\end{equation}
where  $z_{\mathfrak{i}}^2=a^2{\phi'_\mathfrak{i}}^2/\left( 2\mathcal{H}^2f_{\mathfrak{i},{nn}}\right)~$,  $\zeta_{\mathfrak{i}}=\psi+\mathcal{H}\delta\phi_\mathfrak{i}/{\phi_\mathfrak{i}'}$ is Mukhanov-Sasaki variable, $c_{s\mathfrak{i}}^2=n_\mathfrak{i} f_{\mathfrak{i},nn}/f_{\mathfrak{i},n}$ is sound speed square and 
\begin{small}
	\begin{equation}
		\begin{aligned}
			M_1^2&=-\frac{a^2n_2{f_{2{,n}}}^2}{2\mathcal{H}^2}+\frac{a^4n_1n_2f_{1{,n}}f_{2{,n}}}{4\mathcal{H}^2}\left(\frac{c_{s1}^2}{c_{s2}^2}-3\right)+\frac{3}{2}a^2n_2f_{2{,n}}\left(c_{s1}^2+c_{s2}^2\right) -\frac{3a^2n_1n_2f_{2{,n}}f_{1{,nnn}}}{2f_{1{,nn}}},\\
			M_2^2&=-\frac{a^2n_1{f_{1{,n}}}^2}{2\mathcal{H}^2}+\frac{a^4n_1n_2f_{1{,n}}f_{2{,n}}}{4\mathcal{H}^2}\left(\frac{c_{s2}^2}{c_{s1}^2}-3\right)+\frac{3}{2}a^2n_1f_{1{,n}}\left(c_{s1}^2+c_{s2}^2\right) -\frac{3a^2n_1n_2f_{1{,n}}f_{2{,nnn}}}{2f_{2{,nn}}},\\
			M_3&=\frac{a^6n_1n_2f_{1{,n}}f_{2{,n}}}{2\mathcal{H}^2}\left(3-\frac{a^2n_2f_{2{,n}}}{2\mathcal{H}^2c_{s1}^2}-\frac{a^2n_1f_{1{,n}}}{2\mathcal{H}^2c_{s2}^2}\right),~M_4=-\frac{a^6n_1f_{1{,n}}f_{2{,n}}^2}{2\mathcal{H}^3f_{2{,nn}}},~M_5=-\frac{a^6n_2f_{2{,n}}f_{1{,n}}^2}{2\mathcal{H}^3f_{1{,nn}}}.
		\end{aligned}
	\end{equation}
\end{small}

As noted above, the quintom behavior necessitates a sign change in $f_{{1,n}}$, which implies that $c_{s1}^2$ inevitably becomes negative once the EoS parameter of dark energy crosses $w=-1$. Such a negative squared sound speed leads to unavoidable gradient instabilities, which in turn cause the system to be physically unstable. Furthermore, as mentioned before, the quintom-B scenario requires $f_{1{,nn}}<0$, leading to a negative $z_1^2$. This means the dark energy perturbation $\zeta_1$ is a ghost mode because it has a kinetic term with wrong sign. The existence of ghost mode makes the Hamiltonian unbounded from below, leading to severe problems in quantization.  

Consequently, dark energy exhibiting quintom behavior cannot be described successfully by a single perfect fluid minimally coupled to Einstein gravity due to the unavoidable gradient instability around the crossing. In the case of quintom-B built in this way, the situation is worse, because it suffers from both the gradient and ghost instabilities. 

We now turn to the investigation of the dark energy modeled by a single kessence-like scalar field. The most general action for such a model, whose Lagrangian depends  arbitrarily on the field $\phi$ and its first derivative, takes the form:
\begin{equation}\label{grde}
	S_{\phi}=\int d^4x \, \sqrt{-g} \, \mathcal{L}\left(X,\phi\right),
\end{equation}
where $X=-\nabla_\mu\phi\nabla^\mu\phi/2$ is non-negative. By varying the action with respect to the metric, one can directly derive the energy-momentum tensor, which has the form of perfect fluid with the pressure $P=\mathcal{L}(X,\phi)$, energy density $\rho=2XP_{,X}-P$ and the four-velocity $U_\mu=\nabla_{\mu}\phi/\sqrt{2X}$, 
where $``{,X}\,"$ denotes a partial derivative with respect to $X$. Hence the EoS takes the form
\begin{equation}\label{wfield}
	w=\frac{P}{\rho}=-1+\frac{2XP_{,X}}{\rho}.
\end{equation}
For $w$ to cross $-1$, it is necessary for $P_{,X}$ to change sign on the two sides of the crossing point, since $X$ cannot be negative.

Additionally, the possibility of realizing a quintom scenario must be excluded when the Lagrangian density depends only on $X$: $\mathcal{L}=\mathcal{L}(X)$. In this case, the model possesses an exact shift symmetry, $\phi\rightarrow \phi+{\rm constant}$. The corresponding conservation law for the Noether charge is $\mathring{\nabla}_{\mu}j^{\mu}=0$    with $j^\mu=P_{,X}\nabla^\mu\phi=[(\rho+P)/\sqrt{2X}]U^\mu$. One can read from this current the charge density $n=(\rho+P)/\sqrt{2X}$, which scales as $n\propto a^{-3}$ in the expanding universe, just the same as the particle number density in the fluid case. So that $n\propto \rho+P$ cannot cross the zero point, this means $w$ cannot cross the boundary $-1$. Therefore, to have a quintom model the scalar field $\phi$ must appear explicitly in the Lagrangian, i.e., $\mathcal{L}=\mathcal{L}(X, \phi)$.

The total action considered here retains the same structure as that in (\ref{stgr}), except that the dark energy action is now given by (\ref{grde}). Applying a similar treatment, one can derive the quadratic action for scalar perturbations:
\begin{equation}\label{stgr3}
	\begin{aligned}
		S^{(2)}&=\int d^4x  \Big[z_1^2\left({\zeta_1'}^2-c_{s1}^2\partial_i\zeta_1\partial_i\zeta_1+M_1^2\zeta_1^2\right)+z_2^2\left({\zeta_2'}^2-c_{s2}^2\partial_i\zeta_2\partial_i\zeta_2+M_2^2\zeta_2^2\right)\\
		&+M_3\zeta_1\zeta_2+M_4\zeta_1\zeta_2'+M_5\zeta_1'\zeta_2\Big] 
	\end{aligned}
\end{equation}
where $\zeta_2,~z_2^2$ and $c_{s2}^2$ take the same form as those in (\ref{stgr2}), $z_1^2=a^2{\phi'_1}^2\rho_{,X}/(2\mathcal{H}^2)$,  $c_{s1}^2=P_{,X}/\rho_{,X}$ and
\begin{small}
	\begin{equation}
		\begin{aligned}
			M_1^2&=-\frac{a^4n_2^2f_{2,n}^2}{2\mathcal{H}^2}+\frac{a^2n_2f_{2,n}P_{,X}{\phi'}^2}{4\mathcal{H}^2}\left(\frac{c_{s1}^2}{c_{s2}^2}-3\right)+\frac{3}{2}a^2n_2f_{2,n}\left(c_{s1}^2+c_{s2}^2\right) +\frac{a^4n_2f_{2,n}\rho_{{,\phi_1}}}{\mathcal{H}\phi_1'\rho_{,X}}-\frac{a^2n_2f_{2,n}(\rho_{,X})'}{2\mathcal{H}\rho_{,X}} \\
			M_2^2&=-\frac{P_{,X}^2{\phi_1'}^4}{2\mathcal{H}^2}+\frac{a^2n_2f_{2,n}P_{,X}{\phi_1'}^2}{4\mathcal{H}^2}\left(\frac{c_{s2}^2}{c_{s1}^2}-3\right)+\frac{P_{,X}{\phi_1'}^2}{2}\left(3c_{s2}^2+1\right) -\frac{a^2P_{,\phi_{1}}\phi_1'}{2\mathcal{H}}-\frac{P_{,X}\phi_1'\phi_1''}{2\mathcal{H}}-\frac{3P_{,X}{\phi_1'}^2n_2f_{2,nnn}}{2f_{2,nn}}\\
			M_3&=\frac{a^4n_2f_{2,n}P_{,X}{\phi'_1}^2}{2\mathcal{H}^2}\left(3-\frac{a^2n_2f_{2,n}}{2\mathcal{H}^2c_{s1}^2}-\frac{P_{,X}{\phi_1'}^2}{2\mathcal{H}^2c_{s2}^2}\right),~~M_4=-\frac{a^4P_{,X}f_{2,n}^2{\phi_1'}^2}{2\mathcal{H}^3f_{2,nn}},~~M_5=-\frac{a^4n_2f_{2,n}{\phi_1'}^2\rho_{,X}}{2\mathcal{H}^3}.
		\end{aligned}
	\end{equation}
\end{small} 

Given that the $z_1^2$ is proportional to $\rho_{,X}$,  the absence of ghost instability requires $\rho_{,X}$ to remain positive during the evolution. As mentioned, however, a $w=-1$ crossing needs the sign changes in $P_{,X}$, which leads to a negative squared sound speed $c_{s1}^2$ in some region and consequently induces the gradient instability. Therefore, this model inevitably suffers from the gradient instability, which excludes the possibility of a viable quintom scenario in minimally coupled kessence-like scalar field.  

\section{The dark energy in NYTG}\label{III}

The review in the previous section told us that, in Einstein gravity with minimal couplings, the unavoidable instabilities in perturbations prevent either a single perfect fluid or a single kessence-like scalar field from realizing the quintom scenario. In order to address this theoretical problem, we propose a solution within the framework of teleparallel gravity. In our method, the quintom dark energy is still represented by a single perfect fluid or a single kessence-like scalar field. The key point is that we will introduce a coupling of the dark energy to the Nieh-Yan density.  In this way, we incorporated the dark energy into the NYTG model, and as will be shown below, the original instabilities in dark energy perturbations are eliminated in this model. 

In the teleparallel gravity, the gravitation is manifested by spacetime torsion \cite{Aldrovandi:2013wha,Bahamonde:2021gfp}:
\begin{equation}
	T^\rho_{\phantom{1}\mu\nu}=2\Gamma^\rho_{\phantom{1}[\mu\nu]}=2e_A^{\phantom{1}\rho}\left(\partial_{[\mu} e^A_{\phantom{1}\nu]}+\omega^A_{\phantom{1}B[\mu}e^B_{\phantom{1}\nu]}\right),
\end{equation}
where $A,B=0,1,2,3$ are local space tensor indices. Here  $e^A_{\phantom{1}\mu}$ is the tetrad and $\omega^A_{\phantom{1}B\mu}$ is the spin connection. The spacetime metric is obtained from the tetrad via  $g_{\mu\nu}=\eta_{AB}e^A_{\phantom{1}\mu}e^B_{\phantom{1}\nu}$, with $\eta_{AB}$ the Minkowski metric. Under the teleparallel constraints $R^\rho_{\phantom{1}\sigma\mu\nu}=0$ and $\nabla_\rho g_{\mu\nu}=0$, the spin connection can be expressed as $\omega^A_{\phantom{1}B\mu}={\left(\Lambda^{-1}\right)}^A_{\phantom{1}C}\partial_\mu\Lambda^C_{\phantom{1}B}$, where $\Lambda^A_{\phantom{1}B}$ is an arbitrary position dependent Lorentz matrix. Therefore, the theory can be fundamentally described by the language of tetrad $e^A_{\phantom{1}\mu}$ and the Lorentz matrix $\Lambda^A_{\phantom{1}B}$.

As previously mentioned, our model extends GR equivalent teleparallel gravity slightly by considering the coupling of dark energy to the Nieh–Yan density defined as $T_{\lambda\mu\nu}\tilde{T}^{\lambda\mu\nu}=(1/2)\varepsilon^{\mu\nu\rho\sigma} T^\lambda_{\phantom{1}\mu\nu}T_{\lambda\rho\sigma}\,$. 
Including dark energy and other matter components, the action under consideration is:
\begin{equation}\label{anytg}
	\begin{aligned}
		S_{NYTG}&=-\int d^4 x \, \| e\|\left(\frac{\mathbb{T}}{2}+\frac{c}{2}\,\phi\, T_{\lambda\mu\nu}\tilde{T}^{\lambda\mu\nu}\right)+S_{_{DE}}+S_m\\
		&=-\int d^4 x \, \| e\|\left[\left( \frac{1}{8}\,T_{\rho \mu \nu}T^{\rho \mu \nu}+\frac{1}{4}\,T_{\rho \mu \nu}T^{\nu \mu \rho}-\frac{1}{2}\,T_\mu T^\mu\right)+\frac{c}{4}\phi\,\varepsilon^{\mu\nu\rho\sigma} T^\lambda_{\phantom{1}\mu\nu}T_{\lambda\rho\sigma}\right] +S_{_{DE}}+S_m ,
	\end{aligned}
\end{equation}
where $\phi$ is the scalar field corresponding to dark energy, $c$ is a coupling constant and  $ \| e\|=\sqrt{-g}$ is the determinate of the tetrad. The action for dark energy is denoted by $S_{_{DE}}$, while other matter with the action $S_m$ is assumed to be coupled to spacetime minimally through the tetrad. 

Similar to the Chern-Simons term studied frequently in the literature, the Nieh-Yan density is also topological density and appears as a total derivative of the tetrad. The coupling $\phi\, T_{\lambda\mu\nu}\tilde{T}^{\lambda\mu\nu}$ is invariant under the shift of the field $\phi\rightarrow \phi+\phi_0$
by a constant $\phi_0$, up to a surface term. Such kind of couplings are familiar for the axion-like fields. The axion-like fields are pseudo-Goldstone bosons and can only have the couplings preserving the shift symmetry. In this sense, we may treat the dark energy field $\phi$ as axion-like, so that its coupling to the Nieh-Yan density is highly expected within the framework of teleparallel gravity, very similar to the interactions of QCD axion with gauge fields. 

The variations of the action (\ref{anytg}) with respect to $e^A_{\phantom{1}\mu}$ and $\Lambda^A_{\phantom{1}B}$ lead to the following  equations of motion:
\begin{align}
	\mathring{G}^{\mu\nu}+N^{\mu\nu}&=T_m^{\mu\nu}+T_{_{DE}}^{\mu\nu},\label{eomG}\\
	N^{[\mu\nu]}&=0,\label{eomN}
\end{align}
where $\mathring{G}^{\mu\nu}=\mathring{R}^{\mu\nu}-g^{\mu\nu}\mathring{R}/2$ is the Einstein Tensor, $N^{\mu\nu}=c\,\varepsilon^{\mu\lambda\rho\sigma}\partial_\lambda\phi T^\nu_{\phantom{1}\rho\sigma}$, $T_m^{\mu\nu}=(2/\sqrt{-g})(\delta S_m/\delta g_{\mu\nu})$ is the energy-momentum tensor for the matter and $T_{_{DE}}^{\mu\nu}$ is the energy-momentum tensor for the dark energy. Similar to most modified teleparallel gravity models, Eq.(\ref{eomN}), obtained by the variation of $\Lambda^{A}_{\phantom{1}B}$, is not independent of Eq.(\ref{eomG}); it corresponds to the antisymmetric part of that equation. As shown in Ref.\cite{Li:2021wij}, any change caused by $\delta\Lambda^A_{\phantom{1}B}$ can be equivalent to the change cause by $\delta e^A_{\phantom{1}\mu}$ due to the local Lorentz invariance of the action. Owing to this local Lorentz invariance, one may choose the Weitzenb\"{o}ck Gauge in which  $\Lambda^A_{\phantom{1}B}=\delta^A_B$ and $\omega^A_{\phantom{1}B\mu}=0$. In the following sections, we will always adopt this gauge to simplify the calculations. Moreover, there exists an additional equation of motion that follows from the diffeomorphism invariance of the action (\ref{anytg}):
\begin{equation}\label{eomde}
	\mathring{\nabla}_\mu T_{_{DE}}^{\mu\nu}=\mathring{\nabla}_\mu N^{\mu\nu}.
\end{equation}
And it can be easily checked that all the equation of motion given above are in agreement with the the Bianchi identity $\mathring{\nabla}_\mu\mathring{G}^{\mu\nu}=0$ and the conservation law $\mathring{\nabla}_\mu T_m^{\mu\nu}=0$.

\section{The single fluid dark energy in NYTG}\label{IV}
\subsection{background evolution}
In this section, the actions of dark energy and other matter in (\ref{anytg}) are modeled by single perfect fluid which takes the same form as Eq.(\ref{sfluid2}). Therefore the full action (\ref{anytg}) now becomes:
\begin{equation}\label{nydefluid} 
	S_{NYTG}=-\int d^4 x \, \| e\|\,\left[ \frac{\mathbb{T}}{2}+\frac{c}{2} \,\phi_1 T_{\lambda\mu\nu}\tilde{T}^{\lambda\mu\nu}+j_1^\mu a_{1\mu}+f_1\left(\sqrt{-j_1^\mu j_{1\mu}}\right)\right]+S_m(j_2,a_2), 
\end{equation}
where $a_{\mathfrak{i}\mu}=\partial_\mu\phi_\mathfrak{i}+\alpha_\mathfrak{i}\partial_\mu\beta_\mathfrak{i},~ j^\mu_\mathfrak{i}=n_\mathfrak{i}U_\mathfrak{i}^\mu$ and remember that $\mathfrak{i}=1,2$ represent dark energy and matter sector, respectively. Now we apply our model to cosmology. The spacetime under consideration is a spatially flat FRW universe, in which the tetrad can be chosen as:
\begin{equation}\label{frwtetrad}
	e^a_{\phantom{1}\mu}=\mathrm{diag} \{a(\eta),a(\eta),a(\eta),a(\eta)\} .
\end{equation}
Therefore, the metric takes the same form as that given in Eq.(\ref{frwmetric}). Following the substitution of the FRW metric into (\ref{eomG}) and (\ref{eomN}), one can directly obtain the background equations:
\begin{equation}\label{beq1}
	3 \mathcal{H}^2= a^2\left( \rho_m+\rho_{_{DE}}\right),\quad
	-2\mathcal{H}'-\mathcal{H}^2=a^2 \left( P_m+P_{_{DE}}\right),
\end{equation}	
where $\rho_{_{DE}}=f_1(n_1)$ is the dark energy density and $P_{_{DE}}=n_1 f_{1,n}-f_1(n_1)$ is the corresponding pressure. The energy density and pressure of matter take the same form, with the subscripts replaced by $``\,2\,"$. The EoS parameter of single fluid dark energy is identical to that in Eq.(\ref{wfluid}), and the condition $w_{_{DE}}=-1$ is corresponding to $\rho_{_{DE}}+P_{_{DE}}=0$, i.e., $f_{1,n}=0$. Hence, $f_{1,n}$ changes the sign after the crossing. 

Varying the action (\ref{nydefluid}) with respect to $\phi_1,~\alpha_1,~\beta_1$ and $j_1^\mu$ leads to the following equations:
\begin{equation}\label{eomnyfluid} 
	\mathring{\nabla}_\mu \left(n_1 U_1^\mu\right) =\frac{c}{2}\, T_{\lambda\mu\nu}\tilde{T}^{\lambda\mu\nu},\quad U_1^\mu \partial_\mu \alpha_1=-\frac{c \alpha_1}{2 n_1} T_{\lambda\mu\nu}\tilde{T}^{\lambda\mu\nu},\quad U_1^\mu \partial_\mu \beta_1=0,\quad a_{1\mu}=f_{1,n} U_{1\mu}. 
\end{equation}
It can be shown that the above equations are jointly equivalent to Eq.(\ref{eomde}). In the FRW universe, the background components of the four-velocity for a perfect fluid is given by $U^\mu=(1/a,0,0,0)$. Substituting the FRW metric together with this four-velocity into Eq.(\ref{eomde}) yields the background evolution of dark energy:
\begin{equation}\label{beq2}
	n_1'+3\mathcal{H}n_1=0,\quad \alpha_1'=0,\quad \beta_1'=0,\quad \phi_1'=-a \, f_{1,n}.
\end{equation}
It is straightforward to verify that the coupling term in this model has no effect on the background dynamics. The equations in (\ref{beq2}) imply that $\alpha_1$ and $\beta_1$ are functions of spatial position only, i.e., $\alpha_1=\alpha_1(\vec{x}),~\beta_1=\beta_1(\vec{x})$. The same result holds for $\alpha_2$ and $\beta_2$. Consequently, there is a complete freedom in choosing $\alpha_1,~\alpha_2,~\beta_1$ and $\beta_2$, as any choice leads to the same physical background. This freedom can be used to simplify the analysis of cosmological perturbations. For simplicity, we adopt $\alpha_\mathfrak{i}(\vec{x})=\beta_\mathfrak{i}(\vec{x})=0$ when analyzing scalar perturbations in this model.

\subsection{cosmological perturbation analysis}

\subsubsection{Linear perturbation equations for scalar perturbation}

We now turn to the analysis of cosmological perturbations in our model. The scalar  perturbations of tetrad is parameterized as follows: \cite{Li:2022mti,Izumi:2012qj}:
\begin{equation}\label{eperturb}
		e^0_{\phantom{1}0}=a\left(1+A\right),~~e^0_{\phantom{1}i}=a\,\partial_i\chi,~~e^i_{\phantom{1}0}=a\,\partial_i\gamma,~~e^i_{\phantom{1}j}=a\left[\left(1-\psi\right)\delta_{ij}+\partial_i\partial_j\kappa+\epsilon_{ijk}\partial_k\lambda\right].
\end{equation} 
The resulting metric perturbations have the same form as those in (\ref{pertgmn}), with $B=\gamma-\chi$.  Meanwhile, the perturbations of the perfect fluid take the same form as (\ref{pertfluid}). Despite the adoption of the Weitzenb\"{o}ck gauge, the diffeomorphism invariance is still preserved, allowing us to make further gauge choices. The diffeomorphism transformation generated by an infinite small vector $\xi^\mu=(\xi^0,\xi_i+\partial_i\xi)$ is \cite{Li:2021wij}: 
\begin{equation}
	\begin{aligned}
		&A\rightarrow A-\xi^{\prime0}-\mathcal{H}\xi^{0},~\psi\rightarrow\psi+\mathcal{H}\xi^{0}\,,~\chi\rightarrow\chi-\xi^{0}\,,~\gamma\rightarrow\gamma-
		\xi'\,,~\kappa\rightarrow\kappa-\xi\,,\\
		&\lambda\rightarrow\lambda\,,~\delta\phi\rightarrow\delta\phi-\phi'\xi^0\,,~\delta\alpha\rightarrow\delta\alpha\,,~\delta\beta\rightarrow\delta\beta\,,~\delta n\rightarrow\delta n-n'\xi^0\,,~v\rightarrow v+\xi^0\,.
	\end{aligned}
\end{equation} 
Based on the diffeomorphism transformations introduced above, one can define gauge invariant variables that will be used below:
\begin{equation}
	\begin{aligned}
		&\Psi=\psi+\mathcal{H}(\kappa'-B),\,~~~~\Phi=A-\mathcal{H}(\kappa'-B)-(\kappa'-B)',\,~~~\lambda^I=\lambda,\,~~~\delta\alpha_\mathfrak{i}^I=\delta\alpha_\mathfrak{i},\\
		&\delta\beta_\mathfrak{i}^I=\delta\beta_\mathfrak{i},\,~~\delta n^I_\mathfrak{i}=\delta n_\mathfrak{i}-{n}_\mathfrak{i}'(\kappa'-B),~~~\delta\phi^I_\mathfrak{i}=\delta\phi_\mathfrak{i}-\phi'_\mathfrak{i}(\kappa'-B),~~~v^I_\mathfrak{i}=v_\mathfrak{i}+\kappa'-B,
	\end{aligned}
\end{equation}
where $\delta n_\mathfrak{i}$ corresponds to perturbations in the energy density and pressure through $\delta \rho_\mathfrak{i}=f_{\mathfrak{i},n}\delta n_\mathfrak{i}$ and $\delta P_\mathfrak{i}=n_\mathfrak{i} f_{\mathfrak{i},nn}\delta n_\mathfrak{i}$. Linear perturbation equations in this model are obtained by substituting the above parametrizations into Eq.(\ref{eomG}), Eq.(\ref{eomN}) and Eq.(\ref{eomnyfluid}). For simplicity, we adopt the conformal Newtonian gauge, $B=\kappa=0$. The perturbed equation arising from the constraint (\ref{eomN}) can then be written in gauge invariant form as:
\begin{equation}\label{cny}
	\mathcal{H}\delta\phi_1^I+\phi_1'\Psi=0.
\end{equation}
This equation introduces an additional constraint, implying that $\delta\phi_1^I$ is not an independent field but is determined by the constraint. As will be shown later, this constraint can be used to remove the dark energy perturbation from the menu of dynamical degrees of freedom, and also to avoid the original instabilities in dark energy perturbations.

We next turn to the perturbation equations corresponding to Eq.(\ref{eomG}). After straightforward calculations, the full set
of scalar perturbation equations derived from Eq.(\ref{eomG}) is presented below:
\begin{equation}\label{pertg}
	\begin{aligned}
		&6\mathcal{H}(\mathcal{H}\Phi+\Psi')-2\partial_i\partial_i\Psi=-a^2\delta\rho^I_1-a^2\delta \rho^I_2\,,\\
		&2\Psi'+2\mathcal{H}\Phi=-a^2(1+w_1)\rho_1 \, v^I_1-a^2(1+w_2)\rho_2 \, v^I_2\,,\\
		&2\Psi''+4\mathcal{H}\Psi'+2\mathcal{H}\Phi'+2(\mathcal{H}^2+2\mathcal{H}')\Phi=a^2\delta P^I_1+a^2\delta P^I_2\,,\\
		&\Phi-\Psi=2c\phi_1'\lambda^I\,.
	\end{aligned}
\end{equation}
It can be directly seen that the only difference between this model and GR appears in the last equation. The variable $\lambda^I$, which behaves like a viscosity, produces a difference between the perturbations $\Phi$ and $\Psi$. Finally, the perturbation equations given by Eq.(\ref{eomnyfluid}) are
\begin{equation}\label{pertf}
	\begin{aligned}
		&{\delta n_1^{I\,\prime}}+3 \mathcal{H} \delta n_1+n_1\partial_i\partial_iv^I-3n_1\Psi'-\frac{4c\mathcal{H}\partial_i\partial_i\lambda^I}{a}=0,\,\quad \delta\alpha_1^{I\,\prime}=0,\\
		&\delta\phi^{I\,\prime}+a f_{1,n} \Phi+a f_{1,nn} \, \delta n_1^I=0,~\quad \delta\phi^I_1-a f_{1,n} v^I_1=0,\,\,\quad \delta \beta_1^{I\,\prime}=0.
	\end{aligned}
\end{equation}
As mentioned, the equations in (\ref{eomnyfluid}) are jointly equivalent to Eq.(\ref{eomde}), and it can be easily shown that the corresponding perturbation equations in (\ref{pertf}) coincide with those derived from  Eq.(\ref{eomde}). Using the equations above, $\delta n_1^I$ and $v^I_1$ can be expressed in terms of in terms of $\delta\phi^I_1$ and $\Phi$. At first sight, the expression for $v^I_1$ appears to diverge when the dark energy EoS parameter crosses $-1$, due to the vanishing of $f_{1,n}$. However, with the aid of the constraint (\ref{cny}), $v^I_1$ and $\delta n_1^I$ can be rewritten as:
\begin{equation}
	\delta n_1^I=\frac{f_{1,n}}{f_{1,nn}\mathcal{H}^2}\left[-\mathcal{H}^2\Phi+\left(3c_{s1}^2-1\right)\mathcal{H}^2\Psi+\mathcal{H}'\Psi-\mathcal{H}\Psi' \right],\quad v_1^I=\frac{\Psi}{\mathcal{H}}.
\end{equation}
As can be seen from equations in (\ref{pertg}), $\Psi$ and $\Phi$ remain finite when EoS parameter $w_1$ crosses $-1$. As a result, no instabilities arise in the dark energy perturbation as the EoS crosses the cosmological constant boundary.

\subsubsection{Quadratic action for scalar perturbations}

To demonstrate more explicitly that the dark energy perturbations are free from instabilities, we will analyze the quadratic action for scalar perturbations in this model. For our purposes, it is convenient to choose the unitary gauge, $\delta \phi_2=\kappa=0$, such that $\zeta_2=\psi$. Before deriving the quadratic action, we briefly compare our model (\ref{nydefluid}) with the model presented in (\ref{stgr}), where dark energy is described by (\ref{sfluid2}). Given that the determinant of the tetrad obeys  $\|e\|=\sqrt{-g}$, together with the identity $-\mathring{R}=\mathbb{T}+2\mathring{\nabla}_\mu T^\mu$, one finds that the only difference between two actions comes from the coupling term:
\begin{equation}\label{sny}
	S_{NY}=- \int d^4 x \, \|e\| \, \frac{c}{4}\,\phi_1\,\varepsilon^{\mu\nu\rho\sigma} T^\lambda_{\phantom{1}\mu\nu}T_{\lambda\rho\sigma}.
\end{equation}
Substituting the linear perturbations of tetrad (\ref{eperturb}) together with the dark energy perturbations into $S_{NY}$, the quadratic action takes the form: 
\begin{equation}\label{s2ny}
	S^{(2)}_{NY}=-\int d^4 x \, a^2 \left[ 4\, c \, \partial_i\partial_i \lambda \left(\mathcal{H}\delta\phi_1+\psi \phi_1' \right)\right].  
\end{equation}
Because the remaining part of the action (\ref{nydefluid}) is identical to that of the model (\ref{stgr}), the variable $\lambda$, which is purely a tetrad perturbation and does not appear in the metric, contributes only to the quadratic action $S^{(2)}_{NY}$. As a result, varying the action $S^{(2)}_{NY}$ with respect to $\lambda$ leads to the unique constraint:
\begin{equation}\label{s2cny}
	\mathcal{H}\delta\phi_1+\psi \phi_1'=0.
\end{equation}
The equation above is equivalent to the constraint $\zeta_1=0$. Imposing this constraint, the the gauge-invariant variable $\zeta_1$ is expected to vanish in the quadratic action for scalar perturbations, leading to
\begin{equation}
	S^{(2)}=S^{(2)}\left(\zeta_2',\, \zeta_2\right).
\end{equation}
To confirm this expectation explicitly, we  substitute the linear perturbations of tetrad (\ref{eperturb}), together with the perturbations of dark energy and matter sector, into the model (\ref{nydefluid}), from which the quadratic action for scalar perturbations can be directly obtained:
\begin{equation}\label{nyfluids2}
	\begin{aligned}
		S^{(2)}=&-\int d^4 x \, a^2 \bigg[ 2\partial_iA \,\partial_i\psi-\partial_i\psi\,\partial_i\psi-2\mathcal{H}\partial_iA\partial_iB+4\,c\,\partial_i\partial_i\lambda\left(\mathcal{H}\delta\phi_1+\psi\phi_1'\right)+2\partial_i\partial_iB\psi'+3{\psi'}^2+6\mathcal{H}A\psi'\\
		&+a\delta n_1\delta\phi_1'-3a n_1\psi\delta\phi_1'+a n_1\partial_i v_1\partial_i\delta\phi_1 +a^2 A\left(f_{2,n}\delta n_2+f_{1,n}\delta n_1\right) -\frac{1}{2} a^2 n_2 f_{2,n}\left(\partial_iB+\partial_iv_2\right)^2\\
		&-\frac{1}{2} a^2 n_1 f_{1,n}\left(\partial_iB+\partial_iv_1\right)^2+a^2 A^2 \left(f_1+f_2\right)+\frac{1}{2} a^2 f_{1,nn} \delta n_1^2+\frac{1}{2} a^2 f_{2,nn} \delta n_2^2+ a n_1 \delta\alpha_1\delta\beta_1'+ a n_2 \delta\alpha_2\delta\beta_2'\,\bigg].
	\end{aligned}
\end{equation}
It is clear that the variables $\delta\alpha_1,\;\delta\alpha_2,\;\delta\beta_1$ and $\delta\beta_2$ completely decouple from the remaining fields. Moreover, varying the action (\ref{nyfluids2}) with respect to $\delta\alpha_1$ and $\delta\alpha_2$ implies that $\delta\beta_1$ and $\delta\beta_2$ are time-independent, and hence completely determined by their initial values. Consequently, these variables can be be safely neglected in the following analysis. The variables $A,\;B,\;\delta n_{\mathfrak{i}},\;v_{\mathfrak{i}}$ and $\lambda$  are non-dynamical and give rise to the following constraint equations:
\begin{equation}
	\begin{aligned}
		2a^2 A (f_1+f_2)+2\mathcal{H}\partial_i\partial_iB-2\partial_i\partial_i\psi+a^2 f_{1,n} \delta n_1+a^2 f_{2,n} \delta n_2+6\mathcal{H}\psi'=0,\\
		2\mathcal{H}A+a^2 n_1f_{1,n} (B+v_1)+a^2 n_2 f_{2,n} (B+v_2)+2\psi'=0,\\
		a f_{1,nn} \delta n_1+\delta \phi_1'+af_{1,n}A=0,\\
		f_{2,nn} \delta n_2+f_{2,n}A=0,\\
		\delta \phi_1-a f_{1,n} (B+v_1)=0,\\
		a f_{2,n} (B+v_2)=0,\\
		\mathcal{H}\delta\phi_1+\psi \phi_1'=0.
	\end{aligned}
\end{equation}
After substituting these constraints back into the action (\ref{nyfluids2}), the quadratic action for scalar perturbations can be reduced to the following form:
\begin{equation}\label{a2fluid}
	S^{(2)}=\int d^4x \, z_2^2\left({\zeta_2'}^2-c_{s2}^2\,\partial_i\zeta_2\partial_i\zeta_2+M_2^2\zeta_2^2\right),
\end{equation}
where $z_2^2=a^2{\phi_2'}^2/\left( 2\mathcal{H}^2 f_{2,nn}\right), c_{s2}^2=n_2f_{2,nn}/f_{2,n}$ is the sound speed square and the mass square
\begin{equation}
	M_2^2=-\frac{a^2n_1{f_{1{,n}}}^2}{2\mathcal{H}^2}+\frac{a^4n_1n_2f_{1{,n}}f_{2{,n}}}{4\mathcal{H}^2}\left(\frac{c_{s2}^2}{c_{s1}^2}-3\right)+\frac{3}{2}a^2n_1f_{1{,n}}\left(c_{s1}^2+c_{s2}^2\right) -\frac{3a^2n_1n_2f_{1{,n}}f_{2{,nnn}}}{2f_{2{,nn}}}.
\end{equation}
This result is equivalent to the quadratic action (\ref{stgr2}) with $\zeta_1=0$, thereby confirming the expectation discussed above. It is clear that the only dynamical degree of freedom remaining in the quadratic action (\ref{a2fluid}) comes from the matter sector. 
The coupling of dark energy to the Nieh–Yan density  provides an additional constraint and remove the dark energy perturbation from the menu of dynamical degrees of freedom. As a consequence, the instabilities that typically appear in dark energy perturbation around the $w=-1$ crossing are avoided. Therefore, within this framework, one only needs to focus on the background evolution of a single perfect fluid and construct quintom models with confidence.

\subsection{realization of quintom-B scenario}

The subsequent step is to realize the quintom-B behavior using an appropriate single perfect fluid model. We have assumed $\Omega_{m,0}\approx0.31$ and $h\approx0.69$. The background evolution can be rewritten as follows:
\begin{equation}
	3 H^2= \rho_m+\rho_{_{DE}},\quad
	-2\dot{H}-3H^2= P_m+P_{_{DE}}, \quad \dot{n}_1+3Hn_1=0,
\end{equation}
where the dot denotes a derivative with respect to physical time $t$ and $H=\dot{a}/a$. 
For simplicity, we consider the toy model in which the EoS of dark energy depends on the scale factor exactly follow the Chevallier-Polarski-Linder (CPL) parametrization
\begin{equation}
	w_{_{DE}}=w_0+w_a(1-a),
\end{equation}
where $w_0$ and $w_a$ are two constants determined by observational data. 
As mentioned before, we can obtain the model easily:
\begin{equation}
	\rho_{_{DE}}=f_1(n_1)=f_0 \left( \frac{n_1}{n_0}\right)^{w_0+w_a+1} e^{3w_a[ ( n_0/n_1)^{1/3}-1] },
\end{equation}
where $n_1=n_0a^{-3}$ and $f_0$ is the present dark energy density. 

\begin{figure}[htbp]
	\centering
	\captionsetup{justification=raggedright}
	\includegraphics[scale=0.83]{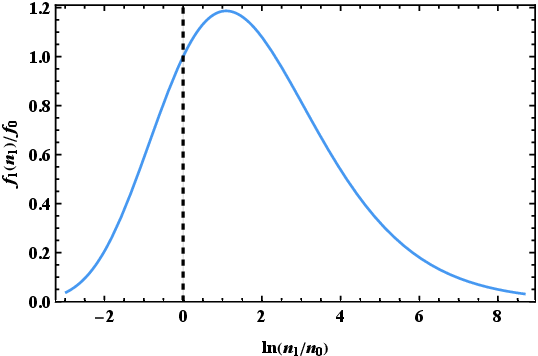}$~~~~$
	\includegraphics[scale=0.85]{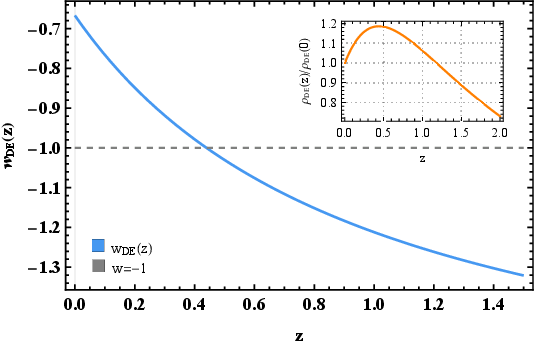}
	\caption{\small{Left panel: The evolution of $f_1(n_1)/f_0$ of single fluid dark energy with $w_0=-0.667,\, w_a=-1.09$. The peak of $f_1(n_1)/f_0$ corresponds to the crossing point, it is very close to the present time. Right panel: The evolution of $w_{_{DE}}(z)$ of single fluid dark energy. The gray dashed line marks the cosmological constant $\Lambda$ with EoS parameter $w=-1$. }}
	\label{ff1}
\end{figure}

The dependence of $f_1(n_1)/f_0$ on $\mathrm{ln}(n_1/n_0)$ and the evolution of $w_{_{DE}}$ as a function of redshift $z$ are presented in FIG.\ref{ff1}, where we adopt $w_0=-0.667,\, w_a=-1.09$ and $f_0\approx7.57\times 10^{-121}$. The values of $w_0$ and $w_a$ are taken from the best-fit results of the observational constraints on the CPL parametrization, as derived in Ref.\cite{DESI:2025zgx} using the combined ``DESI+CMB+Union3" data. As shown in FIG.\ref{ff1}, the function $f_1(n_1)/f_0$ peaks at $n_1/n_0\approx2.99$. As a comparison, the present value $f_1(n_1)/f_0=1$ at $n_1/n_0=1$ is also indicated in the figure. Since the peak is associated with the crossing point, one can evaluate that the dark energy EoS $w_{_{DE}}$ crosses the cosmological constant boundary $w=-1$ at the redshift $z\approx0.44$. 
In addition, this toy model predicts that the dark energy density will decay to zero in the far future.

\section{The single scalar field dark energy in NYTG}\label{V}

\subsection{background evolution}
In this section, the dark energy action appearing in (\ref{anytg}) is presented by a single kessence-like scalar field. The form of the dark energy action is identical to that given in (\ref{grde}), while the matter sector continues to be modeled as a single perfect fluid. Accordingly, the complete action takes the following form:
\begin{equation}\label{nydefield} 
	S_{NYTG}=-\int d^4 x \, \| e\|\,\left[ \frac{\mathbb{T}}{2}+\frac{c}{2} \,\phi_1 T_{\lambda\mu\nu}\tilde{T}^{\lambda\mu\nu}-\mathcal{L}(X,\phi_1)\right]+S_m(j_2,a_2), 
\end{equation}
where $X=-\nabla_\mu\phi_1\nabla^\mu\phi_1/2$ is non-negative. We now turn to the cosmological application of this model and we still consider a spatially flat FRW universe. Upon substituting the tetrad (\ref{frwtetrad}) into the equations of motion (\ref{eomG}), (\ref{eomN}) and (\ref{eomde}), one obtains the corresponding background equations:
\begin{equation}\label{beqfield}
	\begin{aligned}
		3 \mathcal{H}^2= a^2\left( \rho_m+\rho_{_{DE}}\right),\\ -2\mathcal{H}'-\mathcal{H}^2=a^2 \left( P_m+P_{_{DE}}\right),\\	\rho_{,X}\phi_1''+(3P_{,X}-\rho_{,X})\mathcal{H}\phi_1'+a^2 \rho_{,\phi_1} =0,
	\end{aligned}
\end{equation}  
where $\rho_{,\phi_1}=d \rho_{_{DE}}/d\phi_1$, $\rho_{,X}=d \rho_{_{DE}}/dX$ and $P_{,X}=d P_{_{DE}}/d X$. The energy density $\rho$ and pressure $P$ of dark energy and matter are given by:
\begin{equation}
	\rho_m=f_2(n_2)\,,\quad P_m=n_2 f_{2,n}-f_2(n_2)\,,\quad \rho_{_{DE}}=2XP_{,X}-P\,,\quad P_{_{DE}}\equiv P=\mathcal{L}(X,\phi_1)\,.
\end{equation}
In addition, the coupling term is found to have no effect on the background evolution, and the EoS parameter $w_{_{DE}}$ of dark energy takes the same form as that given in Eq.(\ref{wfield}). The condition $w_{_{DE}}=-1$ is equivalent to $\rho_{_{DE}}+P_{_{DE}}=0$,  which occurs when either $\phi_1'=0$ or $P_{,X}=0$. It is important to note, however, that $\phi_1'=0$ is not associated with the crossing. Since $P_{,X}=P_{,X}(X,\phi_1)$ with $X={\phi_1'}^2/2a^2$ is a even function of $\phi_1'$ and $\phi_1'=0$ corresponds to a local extremum, neither $X$ nor $P_{,X}$ changes sign as $\phi_1'$ passes through zero. Therefore, a crossing of $w_{_{DE}}=-1$ necessarily requires a sign change in $P_{,X}$, rather than merely the vanishing of $\phi_1'$.

\subsection{Cosmological perturbation analysis}

\subsubsection{Linear perturbation equations for scalar perturbation}

In this subsection, we investigate cosmological perturbations in the model  (\ref{nydefield}) to determine whether the instabilities in dark energy perturbation can be avoid. The tetrad perturbations take the same form as those in (\ref{eperturb}), and the gauge invariant variables considered in the following analysis are:
\begin{equation}
	\begin{aligned}
		&\Phi=A-\mathcal{H}(\kappa'-B)-(\kappa'-B)',\,~~~~\Psi=\psi+\mathcal{H}(\kappa'-B),\,~~~~\lambda^I=\lambda,\\
		&\delta n^I_2=\delta n_2-n_2'(\kappa'-B),~~~\delta\phi^I_\mathfrak{i}=\delta\phi_\mathfrak{i}-\phi_\mathfrak{i}'(\kappa'-B),~~~v^I_2=v_2+\kappa'-B.
	\end{aligned}
\end{equation}
The linear perturbation equations in this model are therefore obtained by substituting the above parametrizations into the equations of motion derived earlier. For simplicity, we adopt the conformal Newtonian gauge, where $B=\kappa=0$. Then the scalar perturbation equations resulting from Eq.(\ref{eomG}) are listed below:
\begin{equation}\label{pertg2}
	\begin{aligned}
		&6\mathcal{H}(\mathcal{H}\Phi+\Psi')-2\partial_i\partial_i\Psi=-a^2 \rho_{,\phi_1}\,\delta\phi_1^I+\rho_{,X}\phi_1'(\phi_1'\Phi-\delta\phi_1^{I \prime})-a^2\delta \rho^I_2\,,\\
		&2\Psi'+2\mathcal{H}\Phi=P_{,X}\phi_1'\delta\phi_1^I-a^2(1+w_2)\rho_2 \, v^I_2\,,\\
		&2\Psi''+4\mathcal{H}\Psi'+2\mathcal{H}\Phi'+2(\mathcal{H}^2+2\mathcal{H}')\Phi=a^2P_{,\phi_1}\delta\phi_1^I+P_{,X}\phi_1'(\delta\phi_1^{I \prime}-\phi_1'\Phi)+a^2\delta P^I_2\,,\\
		&\Phi-\Psi=2c\phi_1'\lambda^I\,,
	\end{aligned}
\end{equation}
where $\delta \rho_2^I$ and $\delta P_2^I$ are associated with $\delta n_2^I$ through the relation $\delta\rho_2^I=f_{2,n}\delta n_2^I$ and $\delta P_2^I=n_2 f_{2,nn}\delta n_2^I$. These equations are formally identical to those in Eqs.(\ref{pertg}), with the only difference being that the dark energy sector is now associated only with the scalar field $\phi_1$. The variable  $\lambda^I$ still behaves like a viscosity, generating a difference between the perturbations $\Phi$ and $\Psi$. Now we turn to the perturbation equation given by Eq.(\ref{eomN}), which is
\begin{equation}\label{cny2}
	\mathcal{H}\delta\phi_1^I+\phi_1'\Psi=0.
\end{equation}
It is straightforward to verify that this equation coincides exactly with Eq.(\ref{cny}). Because the dark energy sector is entirely described by the scalar field $\phi_1$, its perturbation degree of freedom is completely determined by  $\delta\phi_1^I$. The constraint given in Eq.(\ref{cny2}) indicates that $\delta\phi_1^I$ can be express in terms of the metric perturbation $\Psi$, implying that $\delta\phi_1^I$ does not represent an independent propagating degree of freedom at linear order, when it couples to the tetrad field through the Nieh-Yan density. Furthermore, the equations in (\ref{pertg2}) demonstrate that both metric perturbations $\Psi$ and $\Phi$ remain finite when EoS parameter of dark energy crosses $-1$, i.e., $P_{,X}=0$. Therefore, no instabilities arise in dark energy perturbation at the crossing. 

\subsubsection{Quadratic action for scalar perturbation}

We now proceed to analyze the quadratic action for scalar perturbations in order to demonstrate more explicitly that dark energy perturbation in this model is free from instabilities. For simplicity we choose the unitary gauge, $\delta \phi_2=\kappa=0$, so that $\zeta_2=\psi$. Prior to the derivation, it is worth emphasizing that the only difference between the present model (\ref{nydefield}) and the model considered in (\ref{stgr}), where dark energy is described by (\ref{grde}), originates from the coupling term, with action $S_{NY}$ takes the same form as Eq.(\ref{sny}). As mentioned, the variable $\lambda$ is purely a tetrad perturbation and contributes only to the quadratic action $S_{NY}^{(2)}$. Variation of $S_{NY}^{(2)}$ with respect to  $\lambda$ reproduces the constraint given in Eq.(\ref{s2cny}), which enforces $\zeta_1=0$, a feature also present in single fluid dark energy model. Consequently,  the  gauge-invariant variable $\zeta_1$ is expected to vanish, leaving the perturbations of matter sector as the only remaining dynamical degree of freedom in the quadratic action:
\begin{equation}
	S^{(2)}=S^{(2)}\left(\zeta_2',\, \zeta_2\right). 
\end{equation} 
To confirm this expectation, we substitute the linear perturbations of tetrad (\ref{eperturb}), together with the perturbations of dark energy and matter sector, into the model (\ref{nydefield}). After straightforward calculation, the quadratic action for scalar perturbations has the following form:
\begin{equation}\label{nyfield2}
	\begin{aligned}
		S^{(2)}=&-\int d^4 x \, a^2 \bigg[ 2\partial_iA \,\partial_i\psi-\partial_i\psi\,\partial_i\psi-2\mathcal{H}\partial_iA\partial_iB+4c\,\partial_i\partial_i\lambda\left(\mathcal{H}\delta\phi_1+\psi\phi_1'\right)+2\partial_i\partial_iB\psi'+6\mathcal{H}A\psi'\\
		&+3{\psi'}^2+A\left(a^2 f_{2,n}\delta n_2+\rho_{,X}\phi_1'\delta\phi_1'\right) -\frac{1}{2} a^2 n_2 f_{2,n}\left(\partial_iB+\partial_iv_2\right)^2-\frac{1}{2}\rho_{,X}\delta\phi_1^{\prime\,2}+\frac{1}{2}P_{,X}\partial_i\delta\phi_1\partial_i\delta\phi_1\\
		&+P_{,X}\phi_1'\partial_iB\partial_i\delta\phi_1+P_{,X\phi}\,\phi_1'\delta\phi_1(\phi_1' A- \delta\phi_1')+3P_{,X}\phi_1'\delta\phi_1'\psi-\frac{1}{2}\rho_{,X}{\phi_1'}^2A^2+a^2 A^2 \left(2XP_{,X}-P+f_2\right)\\
		&-a^2P_{,\phi_1}(A-3\psi)\delta\phi_1-\frac{1}{2} a^2 P_{,\phi_1\phi_1}\, \delta \phi_1^2+\frac{1}{2} a^2 f_{2,nn} \delta n_2^2+an_2\delta\alpha_2\delta\beta_2'\,\bigg],
	\end{aligned}
\end{equation}
The variables $\delta\alpha_2$ and $\delta\beta_2$ completely decouple from the other fields. In particular, the general solution for $\delta\beta_2$ is determined by its initial value since action (\ref{nyfield2}) forces it to be time-independent. Therefore, these variables can be safely neglected in the subsequent analysis. In addition, it can be directly found out that $A$, $B$, $\delta n_2$, $v_2$ and $\lambda$ are non-dynamical fields. Variation of the quadratic action (\ref{nyfield2}) with respect to these variables yields the following constraints:
\begin{equation}\label{sconstrian2}
	\begin{aligned}
		2a^2 [X(2P_{,X}-\rho_{,X})-P+f_2]A+a^2\rho_{,\phi_1}\delta\phi_1+2\mathcal{H}\partial_i\partial_iB-2\partial_i\partial_i\psi+a^2 f_{2,n} \delta n_2+\rho_{,X}{\phi_1'}\delta\phi_1'+6\mathcal{H}\psi'=0,\\
		2\mathcal{H}A+a^2 n_2 f_{2,n} (B+v_2)-P_{,X}\phi_1'\delta\phi_1+2\psi'=0,\\
		f_{2,nn} \delta n_2+f_{2,n}A=0,\\
		a f_{2,n} (B+v_2)=0,\\
		\mathcal{H}\delta\phi_1+\psi \phi_1'=0.
	\end{aligned}
\end{equation}
Upon substituting these constraints back into the action (\ref{nyfield2}), the quadratic action for scalar perturbations takes the following form:
\begin{equation}\label{ss22}
	S^{(2)}=\int d^4x~  z_2^2\left(\zeta_2'^2-c_{s2}^2\partial_i\zeta\partial_i\zeta+M_2^2\zeta^2\right),
\end{equation}
where $z_2^2=a^2\phi_2'^2/\left( 2\mathcal{H}^2f_{2,nn}\right)$, $c_{s2}^2=n_2f_{2,nn}/f_{2,n}$ is the sound speed square of matter sector and the mass square is 
\begin{small}
	\begin{equation}
		M_2^2=-\frac{P_{,X}^2{\phi_1'}^4}{2\mathcal{H}^2}+\frac{a^2n_2f_{2_{,n}}P_{,X}{\phi_1'}^2}{4\mathcal{H}^2}\left(\frac{c_{s2}^2}{c_{s1}^2}-3\right)+\frac{P_{,X}{\phi_1'}^2}{2}\left(3c_{s2}^2+1\right) -\frac{a^2\mathcal{L}_{\phi_{_1}}\phi_1'}{2\mathcal{H}}-\frac{P_{,X}\phi_1'\phi_1''}{2\mathcal{H}}-\frac{3P_{,X}{\phi_1'}^2n_2f_{2_{,nnn}}}{2f_{2_{,nn}}}.
	\end{equation}
\end{small}

This result is equivalent to the action (\ref{stgr3}) with $\zeta_1=0$, confirming the expectation mentioned above. As shown by the quadratic action (\ref{ss22}), the only dynamical degree of freedom comes from the matter sector.  The constraint arising from the coupling term eliminate the dark energy perturbation from the menu of dynamical degrees of freedom. As a result, this model circumvents the instabilities in dark energy perturbation, enabling the construction of quintom models by considering only the background evolution of the scalar field.

\subsection{realization of quintom-B scenario}

The next step is to construct a single scalar field dark energy model that is capable of realizing the quintom-B scenario. It is well known that quintessence or phantom models cannot cross the cosmological constant boundary. Hence a more general scalar field Lagrangian must be considered. In this paper, we extended the quintessence model with an additional noncanonical kinetic term. The Lagrangian under consideration is therefore given by:
\begin{equation}
	\mathcal{L}(X,\phi_1)=X+c_1\sqrt{X}-V(\phi_1),
\end{equation}
where $c_1$ is assumed to be non-zero constant. The potential $V(\phi_1)$ is chosen to have the following form:
\begin{equation}\label{vphi}
	V(\phi_1)=\Lambda_{\phi}-\frac{m_{\phi}^2}{2}\phi_1^2+\frac{\lambda_{\phi}}{4}\phi_1^4, 
\end{equation}
where $\Lambda_{\phi}$, $m_\phi$ and $\lambda_\phi$ are chosen to be positive constants. The last term in potential (\ref{vphi}) is included to prevent the model from developing a tachyonic instability, which would arise if the potential were unbounded from below. The background equations can be rewritten into :
\begin{equation}\label{bfieldnum}
	3 H^2= \rho_m+\rho_{_{DE}},\quad
	-2\dot{H}-3H^2= P_m+P_{_{DE}},\quad \ddot{\phi}_1+3H\left(\dot{\phi}_1+\frac{c_1}{\sqrt{2}}\mathcal{P}\right)+V_{\phi_1}=0,
\end{equation}
where the dot denotes a derivative with respect to physical time $t$ and $\mathcal{P}\equiv \mathrm{sgn}(\dot{\phi}_1)\in\left\lbrace+1, -1, 0\right\rbrace $, corresponding to $\dot{\phi}_1>0$, $\dot{\phi}_1<0$ and $\dot{\phi}_1=0$, respectively. The energy density and the pressure of dark energy are now expressed as:
\begin{equation}
	\rho_{_{DE}}=\frac{\dot{\phi}_1^{\, 2}}{2}+V\left(\phi_1\right)  ,\quad P_{_{DE}}=\frac{\dot{\phi}_1^{\, 2}}{2}+\frac{c_1}{\sqrt{2}}|\dot{\phi}_1|-V\left(\phi_1\right).
\end{equation}
Consequently, the EoS parameter of dark energy is 
\begin{equation}\label{eos2}
	w_{_{DE}}=-1+\frac{2\dot{\phi}_1^{\, 2}+\sqrt{2}c_1|\dot{\phi}_1|}{\dot{\phi}_1^{\, 2}+2V(\phi_1)}.
\end{equation}
Since the energy density $\rho_{_{DE}}$ is positive definite during the evolution of $\phi_1$, the denominator of the EoS parameter in Eq.(\ref{eos2}) must remain positive during the evolution. Meanwhile, if $c_1$ is positive, the numerator cannot cross zero. Therefore, a negative $c_1$ is generally required for the crossing. 

\begin{figure}[htbp]
	\centering
	\subfigure{\includegraphics[scale=0.84]{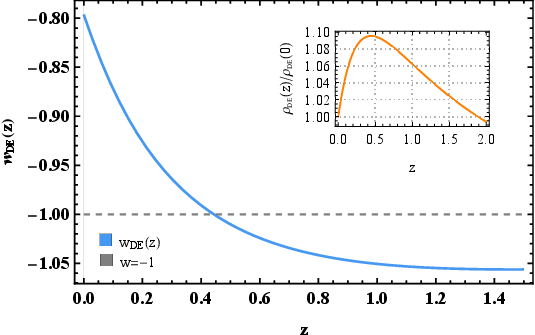}}
	\captionsetup{justification=raggedright}
	\caption{\small{The evolution of EoS parameter $w_{_{DE}}(z)$ as a function of redshift $z$ and the gray dashed line marks the cosmological constant $\Lambda$ with EoS parameter $w=-1$. }}
	\label{wz1}
\end{figure}

We next numerically solve the background equations in (\ref{bfieldnum}). To illustrate the evolution of the EoS parameter $w_{_{DE}}$, we have assumed $\Omega_{m,0}\approx0.31$ and $h\approx0.69$. Within the above parameter ranges, the model parameters can be chosen with considerable flexibility. In particular, the evolution of the EoS $w_{_{DE}}$ as a function of redshift $z$ is presented in FIG.\ref{wz1}, where we adopt the parameter values: $ c_1=-1.14\times10^{-60},~ \Lambda_{\phi}=5.29\times10^{-121},~ m_{\phi}=1.12\times10^{-60}$ and $\lambda_{\phi}=7.69\times10^{-121}\,$. The initial conditions are fixed to $ \dot{\phi}_{1_{ini}}=-8.06\times10^{-61}$ and $\phi_{1_{ini}}=0.6$. In this case, the EoS parameter  $w_{_{DE}}(z)$ is found to evolve smoothly and crosses the cosmological constant boundary at $z\approx0.44$. At the present epoch, i.e., $z=0$, the model yields $w_0\approx-0.8$. We emphasize that the model considered here serves only as an illustrative example, and a detailed comparison with observational data will be pursued in future work.

\section{Parity violation in gravitational waves}\label{VI}

In this section we demonstrate that our model not only avoid the instabilities in dark energy perturbation, but also gives rise to parity violation in gravitational waves. From the linear perturbation equations one can find that the tensor $N^{\mu\nu}$ in Eq.(\ref{eomG}) does not contain any vector perturbation \cite{Li:2020xjt}. Hence the vector sector contains no dynamical degrees of freedom, as in GR. We next consider the tensor sector, where tensor perturbations arise only from the tetrad $e^i_{\phantom{1}j}$:
\begin{equation}
	e^i_{\phantom{1}j}=a\left( \delta_{ij}+\frac{1}{2}h^T_{ij}\right).
\end{equation}
The tensor perturbations $h^{T}_{ij}$ are transverse and traceless, i.e., $\partial_ih^T_{ij}=\delta^{ij}h^T_{ij}=0$. No matter dark energy is a single perfect fluid or single kessence-like scalar field, the quadratic action for tensor perturbations always takes the same form, which coincides with that obtained in Ref.\cite{Li:2020xjt}:
\begin{equation}\label{tensoraction}
	S=\int d^{4}x \ \frac{a^{2}}{8} \left({h^{T}_{ij}}'{h^{T}_{ij}}'-\partial_{k}h^{T}_{ij}\partial_{k}h^{T}_{ij}
	-2 c\phi_1' \epsilon_{ijk}h^{T}_{il}\partial_{j}h^{T}_{kl}\right)\,,
\end{equation}
After expanding the tensor perturbations in Fourier space by circular polarization bases 
\begin{equation}
	h_{ij}^T(t,\vec{x})=\frac{1}{(2\pi)^{3/2}}\int d^3\vec{k}~h_{ij}^T(t,\vec{k}) \, e^{i\vec{k}\cdot\vec{x}}~,~h_{ij}^T(t,\vec{k})=\sum_{A=L,R}h_A(t,\vec{k}) \, \hat{e}^A_{ij}(\vec{k}),
\end{equation}
the quadratic action can be rewritten as
\begin{equation}\label{tensoraction2}
	S=\sum_{A=L,R}\int d\eta d^{3}k \ \frac{a^{2}}{4} \left[ h^{\prime\,2}_A-(k^{2}+2\mathfrak{p}_{A}c \phi_1'k)h^2_A\right]\,,
\end{equation}
where $\mathfrak{p}_L=-1,\ \mathfrak{p}_R=1$, and we mark $h^{\prime *}_Ah'_A\rightarrow h^{\prime2}_A,\ h^*_Ah_A\rightarrow h^2_A$ for simplicity. The equation for tensor perturbations derived from (\ref{tensoraction2}) is \begin{equation}\label{eomh}
	h_A''+2\mathcal{H}h_A'+\omega_A^2h_A=0,
\end{equation}	
where $\omega_A^2=k^2+2\mathfrak{p}_{A}c \phi_1'k$.	From (\ref{eomh}), one can find that gravitational waves with different helicities will have different phase velocities: $v_p^A=\omega_A/k\approx1+\mathfrak{p}_Ac\phi_1'/k\equiv1+\mathfrak{p}_AaM_{PV}/(2k)$. This phenomenon is known as velocity birefringence of gravitational waves and provides an explicit signal of parity violation in this model. This is a unique theoretical prediction of the mechanism proposed in this paper, makes it distinguishable from other models of non-minimal couplings or modified gravities. It has obtained constraints by current experimental data of gravitational waves and will accept more tests in the future. 
Through the full Bayesian inference on the gravitational wave events of binary black hole merges (BBH), an upper bound on the parameters of NYTG model was placed in Ref.\cite{Wu:2021ndf}: $M_{PV}<6.5\times10^{-42}$GeV. In our framework this implies the constraint: $|c \phi_1'/a|<3.25\times10^{-42}$GeV.

\section{Conclusion}\label{VII}

Current observational data favor a dynamical dark energy with quintom-B behavior. However, in Einstein gravity with minimal couplings, the unavoidable instabilities in perturbations prevent either a single perfect fluid or a single kessence-like scalar field from realizing the quintom scenario. In this paper we proposed a possible way out of this difficulty by incorporating the dark energy into the NYTG model, which is a slight modification to the GR equivalent teleparallel gravity. In our model, the quintom dark energy is still represented by a single perfect fluid or a single kessence-like scalar field, but it has a direct coupling to the Nieh-Yan density. Through the analyses on the cosmological perturbations of this model, we showed that the introduced coupling brings an extra constraint and forces the gauge-invariant dark energy perturbation to vanish. In this way, the dark energy perturbation is removed from the menu of the dynamical degrees of freedom and the original instabilities in it are also eliminated. 

This property originates from the fact that the theory of teleparallel gravity with Nieh-Yan modification is effectively a tetrad theory rather than a metric one. Generally the tetrad field has sixteen components, and compared with the metric, the additional six components may alter the constraint structure of the gravity theory. This may in turn impose extra constraint on the dark energy scalar field through the Nieh-Yan coupling between it and the tetrad field. In this paper, we showed that it is the pseudo-scalar perturbation $\lambda$ that constrains the gauge-invariant dark energy perturbation to zero. The pseudo-scalar perturbation comes from the space-space component of the tetrad and appears as a Lagrange multiplier in the quadratic action for linear perturbations. It is absent in the metric decomposition. We would like to point out that we only found this property in the linear perturbation theory around the spatially flat FRW universe. If the background universe is closed or open, the scalar perturbation in this model remains propagating, as demonstrated earlier in Ref.\cite{Li:2021wij}. Furthermore, whether the non-propagating nature of the scalar perturbation around the spatially flat FRW universe is valid beyond the linear perturbation theory remains an open question, addressing it requires a non-linear perturbation analysis.

We also showed that the Nieh-Yan coupling of dark energy leads inevitably parity violations in gravitational waves. This can be considered as unique theoretical prediction of this mechanism, further signals the dynamical nature of dark energy, and is expected to be tested by future gravitational wave experiments. 

In summary, our work provides a new approach to the investigation of quintom dark energy models.
\\

\textbf{Acknowledgments:} This work is supported by NSFC under Grant No. 12575067 and 12247103, and the National Key R\&D Program of China Grant No. 2021YFC2203102. 

\bibliography{nyde}{}
\bibliographystyle{elsarticle-num}

\end{document}